\documentclass[twocolumn]{aastex631}
\usepackage{amsmath}

\newcommand{\psr}{PSR\,J0002+6216}
\newcommand{\ctb}{CTB 1}
\newcommand{\gctb}{G116.9+0.2}

\received{2023-09-25}
\revised{2023-10-25}
\accepted{2023-10-30}
\submitjournal{\apj}

\shorttitle{J0002+6216 Proper Motion}
\shortauthors{Bruzewski et al.}

\begin{document}

\title{Cannonball or Bowling Ball: A Proper Motion and Parallax for PSR J0002+6216}

\author[0000-0001-7887-1912]{S. Bruzewski}
\affiliation{Department of Physics and Astronomy, University of New Mexico, Albuquerque, NM 87131, USA}

\author[0000-0001-6672-128X]{F.K. Schinzel}
\altaffiliation{An Adjunct Professor at the University of New Mexico.}
\affiliation{National Radio Astronomy Observatory, P.O. Box O, Socorro, NM 87801, USA}

\author[0000-0001-6495-7731]{G.B. Taylor}
\affiliation{Department of Physics and Astronomy, University of New Mexico, Albuquerque, NM 87131, USA}

\author{P. Demorest}
\affiliation{National Radio Astronomy Observatory, P.O. Box O, Socorro, NM 87801, USA}

\author{D. A. Frail}
\affiliation{National Radio Astronomy Observatory, P.O. Box O, Socorro, NM 87801, USA}

\author[0000-0002-0893-4073]{M. Kerr}
\affiliation{Space Science Division, US Naval Research Laboratory, Washington, DC 20375, USA}

\author[0000-0003-0101-1986]{P. Kumar}
\affiliation{Department of Physics and Astronomy, University of New Mexico, Albuquerque, NM 87131, USA}

\correspondingauthor{S. Bruzewski}
\email{bruzewskis@unm.edu}

\begin{abstract}
    We report the results of careful astrometric measurements of the Cannonball pulsar J0002+6216 carried out over three years using the High Sensitivity Array (HSA). We significantly refine the proper motion to $\mu=35.3\pm0.6$ mas yr$^{-1}$ and place new constraints on the distance, with the overall effect of lowering the velocity and increasing the inferred age to $47.60\pm0.80$ kyr. Although the pulsar is brought more in line with the standard natal kick distribution, this new velocity has implications for the morphology of the pulsar wind nebula that surrounds it, the density of the interstellar medium through which it travels, and the age of the supernova remnant (CTB 1) from which it originates.
\end{abstract}

\section{Introduction} \label{sec: intro}

Identifying compact objects associated with supernova remnants provides a unique window into the diverse outcomes of core-collapse supernovae. A veritable zoo of such young neutron stars have been found \citep{2023Univ....9..273P} including central compact objects (CCOs) and various types of magnetar, such as soft gamma-ray repeaters (SGRs) and anomalous X-ray pulsars (AXPs) \citep{2001ApJ...559..963G}. Young radio pulsars comprise the majority of neutron stars found in supernova remnant (SNR) associations \citep{2019JApA...40...36G, 2012AdSpR..49.1313F}. The study of these systems has helped constrain initial pulsar periods, magnetic fields, and beaming fractions, in addition to their kick velocity at birth \citep{1993ApJ...408..637F, 1997MNRAS.291..569H, 2022MNRAS.514.4606I, 2023MNRAS.519.5893K,2005MNRAS.364.1397J}.

\citet[][henceforth Paper I]{2019ApJ...876L..17S} noted that the 115 ms $\gamma$-ray and radio pulsar \psr\ was found at the head of a bow-shock pulsar wind nebula (PWN). Follow-up radio and X-ray observations resolved the PWN and showed that it was consistent with a high Mach shock formed from the wind from the energetic high velocity \psr\ coming into ram pressure balance with its surrounding medium \citep{2023ApJ...945..129K}. Remarkably, the tail of the PWN extends at least 7-10 arcminutes from \psr, pointing backward toward the geometric center of the Galactic SNR CTB-1 (\gctb) 28 arcminutes away. The authors argued that \ctb\ is the remnant of the supernova that produced \psr, indicating that \psr\ was born with an unusually high velocity ($V_{\rm PSR}>1000$ km s$^{-1}$) that allowed it to escape the parent SNR.

Not all of the young pulsars found in or near SNRs are real associations. Often the case is made on the basis of positional coincidence and some agreement between ages and distances.  The gold standard for making such claims is accurate pulsar proper motion and parallax measurements. Often both the \textit{magnitude} and \textit{direction} of proper motion can reinforce the claimed association by showing that the pulsar likely originated near the geometric center of the SNR \citep{2012ApJ...748L...1D, 2012ApJ...755..151V, 2019ApJ...877...78S, 2021MNRAS.507L..41J, 2022ApJ...932..117L}. In other cases, proper motion measurements either exclude an association or require more complex scenarios to remain tenable \citep{2006ApJ...652..554B, 2021ApJ...908...50D, 2022A&A...659A..41E}. Pulsars (or PWN) with large offsets from the geometric center of the SNR, including pulsars \textit{outside} the SNR \citep[e.g,][]{2012ApJ...746..105N, 2023MNRAS.523.2850M}, are especially important for constraining the high birth velocity tail of PSRs \citep{1994ApJ...437..781F, 2005MNRAS.360..974H, 2017A&A...608A..57V}. The burden of proof is high, since in these cases the probability of a chance association is greatly increased \citep{1995PASA...12...76G}. Proper motion measurements of such associations have given mixed results at best \citep{2008ApJ...674..271Z, 2009ApJ...706.1316H}. In the case of \psr/CTB-1, Paper I bolstered their case by using nearly a decade of data from the \textit{Fermi Gamma-Ray Space Telescope} to infer a proper motion of $\mu=115\pm33$ mas yr$^{-1}$ and $\theta_{\mu}=121^\circ\pm13$. This value, although of modest significance, agrees in magnitude and direction with the PSR-SNR angular offset, assuming an age for the system of 10 kyrs.

In this work, we have undertaken VLBI observations over the course of several years towards providing an accurate measure for the proper motion and parallax of \psr. Having such values allows us to provide a more definitive association between \psr\ and CTB-1, as well as to probe the implications of the SNR age implied from the more accurately measured velocity of the pulsar. Section \ref{sec: data} will discuss our observational setup, data processing, and the fitting of parallax and proper motion to our data. Section \ref{sec: results} describes the best-fit parameters, and Section \ref{sec: discussion} discusses their implications. Finally Section \ref{sec: conclusions} concludes and summarizes this work.

\section{Data} \label{sec: data}

\subsection{Observations} \label{subsec: observations}

Observations were performed under project code BS278 (VLBA/19B-048) with the High Sensitivity Array (HSA\footnote{\url{https://science.nrao.edu/facilities/vlba/HSA}}), comprised of NRAO's Very Long Baseline Array (VLBA) and the phased Karl G. Jansky Very Large Array (VLA), as well as the Effelsberg 100\,m radio telescope. The observations spanned a period of just over two years (2020 February 10 -- 2022 July 30). The antennas unavailable during a specific observation, together with additional summarizing characteristics, are noted in Table \ref{table: observations}. 

Initially, we performed a test observation (BS278Z) to check for suitable phase reference calibrators, both for VLBA-only in-beam calibration (selected from the VLA L-band image presented in Paper I) and nearby bright calibrator sources selected from the VLBA calibrator catalog\footnote{\url{https://obs.vlba.nrao.edu/cst}}. This test did not yield a suitable in-beam calibrator and only identified one suitable phase calibrator, J0003+6307, separated by 0.87$^\circ$ with a flux density of 46.5$\pm$1.3 mJy and a peak of 10.48$\pm$0.24 mJy\,beam$^{-1}$. Given the weak nature of this calibrator, we included a secondary phase calibrator J2339+6010, which is at a distance of 3.53$^\circ$ from our target with a flux density of about 70\,mJy and a peak of about 36 mJy\,beam$^{-1}$. In addition, we observed J0014+6117 at a distance of 1.71$^\circ$ (about 30 mJy / 15 mJy\,beam$^{-1}$ peak) as a VLA phasing calibrator. J0014+6177 can also be used as cross-check calibrator. However, it seems to be resolved out on the longest baselines. J0319+4130 and J0137+3309 served as fringe finders, bandpass calibrators, and flux references. The calibrators and phase centers are summarized in Table \ref{table: cablibrators}.

\begin{deluxetable}{lllll} \label{table: observations}
\tablecaption{HSA Observation Summary}
\tablehead{\colhead{Date of Obs.} & \colhead{S} & \colhead{A} & \colhead{Missing} & \colhead{Issues} }
    \startdata
    2019-07-20\tablenotemark{a} & Z & 11 & EF,Y & --\\
    2019-10-09\tablenotemark{a} & A & 12 & --  & NL,EF gaps\\
    2020-02-10\tablenotemark{a} & B & 11 & MK & EB,Y gaps\\
    2020-05-04 & C & 11 & SC    & KP,FD,OV gaps\\
    2020-08-08 & D & 12 & --    & SC,KP,NL,LA gaps\\
    2020-10-05 & E & 11 & HN    & KP,SC gaps\\
    2020-12-14 & F & 12 & --    & OV high winds\\
    2021-03-28\tablenotemark{a} & G & 11 & OV & SC,PT,MK 1h lost\\
    2021-06-18 & H & 11 & SC    & Y late, FD timing\\
    2021-09-20 & I & 12 & --    & --\\
    2022-02-14 & J & 11 & HN    & MK missing data\\
    2022-07-30 & K & 10 & KP,MK & --\\
    \enddata
\tablecomments{The \textit{S} column denotes the segment of BS278, and the \textit{A} column counts the number of participating antennas. \textit{Issues:} Gaps are noted where a significant amount of data are missing from correlation.}
\tablenotetext{a}{Non-detections}
\end{deluxetable}
\vspace{-\baselineskip}

The Z and A segments of BS278 were observed using the 18 cm receiver and a frequency range of 1.568--1.824\,GHz and the RDBE personality of the VLBA backend with a recording data rate of 2 Gbps. This resulted in non-detection of our target, primarily due to strong radio interference at some of the antenna sites. Subsequent observations (BS278C onward) thus used a lower frequency range centered on a mostly interference-free part of the spectrum, 1.268--1.524\,GHz in the 21\,cm wavelength band. For Effelsberg we utilize the center horn of the 7-beam 21\,cm receiver. In addition, the phased-VLA was used to record the full allowable bandwidth of 1\,GHz (starting with segment C) to determine the pulsar ephemerides using VLA's WIDAR correlator for later binned correlation of the HSA observations. The phased-VLA observations yielded detections in all cases with an average flux density of $\sim 30$\,uJy. The pulsar is linearly polarized and shows a rotation measure of $-178.5\pm2.5$\,rad\,m$^{-2}$. 

The HSA correlation was performed using the VLBA DiFX correlator \citep{2011PASP..123..275D} using 256 correlation channels per one of the eight dual-polarization spectral windows with 32 MHz bandwidth each and a 0.5\,s dump time. This allows full coverage of 256\,MHz in bandwidth centered on 1.396\,GHz. As mentioned before, for each phased VLA observation, pulsar timing data was recorded to obtain pulsar ephemerides of our target pulsar, which were then supplied to DiFX for binned correlation, where 10 bins in pulsar spin phase were formed to increase the signal-to-noise ratio of the pulsar emission, with the first bin (``ON'') centered on the main pulse of the pulsar. The VLBA DiFX correlator is set such that the flux density remains the same as the unbinned equivalent. In addition, multiple phase centers were placed on locations of known bright sources that were identified from the wide-field VLA L-band observation mentioned above. In addition, we updated the target phase center four times throughout the campaign to account for the pulsar proper motion and keep the pulsar close to the correlation phase center. Correlation phase centers are provided in Table \ref{table: cablibrators}.

\begin{deluxetable*}{rllll} \label{table: cablibrators}
\tablecaption{Calibrators and Phase Centers}
\tablehead{\colhead{Name} & \colhead{R.A.} & \colhead{Dec.} & \colhead{Dist.} & \colhead{Description} \\ \colhead{} & \colhead{h:m:s} & \colhead{d:m:s} & \colhead{deg.} & \colhead}
    \startdata
    J0001+6051 & 00:01:07.099891 & +60:51:22.79829 & 1.43 & Position Check Calibrator\\
    J0003+6307 & 00:03:36.511479 & +63:07:55.87126 & 0.87 & Primary Phase Calibrator\\
    J0014+6117 & 00:14:48.792109 & +61:17:43.54262 & 1.71 & VLA phasing calibrator\\
    J0137+3309 & 01:37:41.299543 & +33:09:35.13377 & 3.99 & Position Check Calibrator\\
    J0319+4130 & 03:19:48.160114 & +41:30:42.10568 & -- & Fringe Check / Bandpass Calibrator\\
    J2236+2828 & 22:36:22.470849 & +28:28:57.41328 & -- & Flux Check / Fringe Check\\
    J2339+6010 & 23:39:21.125227 & +60:10:11.84951 & 3.53 & Secondary Phase Calibrator\\
    \tableline
    J0003+6219 & 00:02:53.527 & +62:19:16.940 & -- & Correlation phase center\\
    J0002+6205 & 00:01:44.120 & +62:04:30.690 & -- & Correlation phase center\\
    J0001+6228 & 00:00:43.195377 & +62:27:59.722930 & -- & Correlation phase center\\
    J0000+6222 & 00:00:17.959930 & +62:22:20.034252 & -- & Correlation phase center\\
    J0000+6215 & 00:00:24.397047 & +62:15:21.717124 & -- & Correlation phase center\\
    J0000+622A & 00:00:17.753    & +62:22:14.354 & -- & Correlation phase center\\
    \tableline
    TARGET & 00:02:58.17 & +62:16:09.4 & -- & Correlation phase center for segment B\\
    TARGET & 00:02:58.205 & +62:16:09.510 & -- & Correlation phase center for segments D--E\\
    TARGET & 00:02:58.207 & +62:16:09.500 & -- & Correlation phase center for segments F--H\\
    TARGET & 00:02:58.210 & +62:16:09.4958 & -- & Correlation phase center for segments I--K \\
    \enddata
\tablecomments{List of Calibrators and correlation phase centers for BS278C-K, including updated target positions. Coordinates are in J2000.}
\end{deluxetable*}
\vspace{-2\baselineskip}

\subsection{Calibration \& Model-Fitting} \label{subsec: fitting}

The correlated data were calibrated using the Astronomical Image Processing System \citep[AIPS;][]{2003ASSL..285..109G} using the included \texttt{VLBAUTIL} procedures. Standard astrometric calibration was performed applying corrections for Earth's orientation, the ionosphere, for digital sampling, and delay. After this, bandpass, sampler, gain, and parallactic angle corrections were applied, followed by determination and application of phase corrections using the calibrator, J0003+6307. For applying calibration solutions to the different phase centers and bins the \texttt{VLBAMPHC} procedure was used. After this, the calibration corrections were applied to the correlated data and split by observed source and written to separate UVFITS formatted files for further analysis.

Then we used the Difmap package \citep{1997ASPC..125...77S} for additional manual flagging of radio frequency interference, primarily from Global Navigation Satellite System interference, such as GPS, Galileo, or GLONASS in the 21\,cm wavelength band. We then used $u$,$v$-plane model-fitting of a delta function, as a good representation for an unresolved point source, to determine the position of our target and calibrators for each epoch. The detections were typically at the 5$\sigma$ level, which did not allow for robust fitting of a Gaussian function to the visibilities. We also used Difmap to produce restored clean images for the calibrators and target. 

One issue arising from this approach is that, while Difmap's model fitting excels in this particular case of faint sources, it does not provide much in the way of uncertainties for the calculated best-fit coordinates. Ideally we would like to have an estimate at each epoch for how well the position of a source of a certain flux density can be localized amid some level of background noise. We do this by making a copy of our original visibility data, then using CASA \citep{2022PASP..134k4501C} to replace all data-points with an artificial point source of known position and flux density (set to approximately match the properties of \psr). Random simple noise is then added to the data, which we re-scale using the existing weights. Difmap is then provided this artificial data, as well as a perturbation of the position to use as an initial starting model, and made to perform model fitting for 50 steps, after which the final model is recorded to disk. 

Conveniently, none of the algorithms involved in this simulation are all that intensive, and so the entire pipeline can be run using a single script on a single core. Given that the only difference between two simulations will be the seed of the random noise, the problem is embarrassingly parallelizable, and so we perform 1000 such simulations at each epoch. Once the final models of all simulations have been collected and tabulated, we simply need to characterize the scatter of the models to establish our fitting uncertainty. We note that the scatter in Right Ascension and Declination seem to be highly correlated in all epochs, likely due to our $u$,$v$-coverage, such that we cannot disregard the off-diagonal terms of the covariance matrix at this stage. Instead we opt to record the covariance matrix at each epoch, measured over the coordinates $\Delta\alpha^*=(\alpha-\bar{\alpha})\cdot\cos\bar{\delta}$ and $\Delta\delta = (\delta - \bar\delta)$\footnote{Note that this cosine correction will appear fairly regularly throughout this work}.

\subsection{Atmospheric Effects} \label{subsec: atmo}

For our observations, the positional uncertainties on a given date will be dominated by the effects of the ionosphere. This can be corrected for at some level by ionospheric corrections performed during calibration, but for corrections at the milliarcsecond level, typically further refinements are required. Similar work \citep{2016ApJ...828....8D, 2019ApJ...875..100D,2023MNRAS.519.4982D} has typically made use of one or more relatively bright in-beam calibration sources, which (if truly extra-galactic and stationary) can be simultaneously phase-calibrated, effectively removing atmospheric effects by providing an absolute position offset between the two sources at the sub-milliarcsecond level. 

Unfortunately, such calibration is not possible in the case of J0002+6216. The area immediately surrounding the pulsar is devoid of any other sources out to the edges of the VLBA field of view. Furthermore, as this source is quite faint ($\sim$22--30\,uJy in single dish/phased VLA observations), we require the additional sensitivity from the inclusion of Effelsberg and the phased VLA, which have an even smaller field of view, further ruling out this option. The typical detection in the ``ON'' bin of our observations, resulted in a flux density of $\sim$14 uJy, suggesting that a significant fraction is lost due to residual ionospheric phase errors that dominate in this frequency range.

Lacking a suitable source in the field of view, we opt to leverage a nearby but out-of-beam source. This secondary source, dubbed SRC2, has a peak flux density near 700 uJy\,beam$^{-1}$, meaning its own positional uncertainties (as measured via simulations described in Section \ref{subsec: fitting}) are more than an order of magnitude better constrained than those of our target. Its position seems to experience ionospheric variations on the same scale as other calibrators, and has no apparent proper motion across our observations. As SRC2 is fairly close to our target, we can assume that it lies behind roughly the same region of the ionosphere \citep[as described approximately by][]{1999ASPC..180.....T}, and should have its position perturbed similarly. With this in mind we use the displacement of SRC2 from its median position to apply a correction to the position of J0002+6216 at each epoch. This should very roughly approximate the results of the aforementioned simultaneous phase-calibration, albeit with higher uncertainties. These updated positions are reported in Table \ref{table: positions}.

\begin{deluxetable}{llll} \label{table: positions}
\tablecaption{Target Positions}
\tablehead{\colhead{Date} & \colhead{Position (J2000)} & \colhead{$\sigma_\alpha^*$} & \colhead{$\sigma_\delta$}\\ \colhead{MJD} & \colhead{ } & \colhead{$\mathrm{mas}$} & \colhead{$\mathrm{mas}$}}
\startdata
58973.44 & 00h02m58.20495s +62d16m09.50991s & 1.29 & 1.11 \\
59069.70 & 00h02m58.20652s +62d16m09.50328s & 1.18 & 1.07 \\
59127.89 & 00h02m58.20686s +62d16m09.50229s & 1.19 & 1.09 \\
59197.80 & 00h02m58.20773s +62d16m09.49950s & 1.21 & 1.22 \\
59383.63 & 00h02m58.21090s +62d16m09.49124s & 1.28 & 1.12 \\
59477.80 & 00h02m58.21150s +62d16m09.49155s & 1.17 & 1.10 \\
59624.17 & 00h02m58.21284s +62d16m09.48623s & 1.13 & 1.05 \\
59790.60 & 00h02m58.21578s +62d16m09.47532s & 1.23 & 1.12
\enddata
\tablecomments{Position of J0002+6216 at each epoch it was detected. Uncertainties are derived from simulations of position fitting to our UV-data, as well as systematic uncertainty from the ionosphere. Here we report only the diagonal terms of the covariance matrix as our uncertainties, as the correlation is fairly small.}
\end{deluxetable}

\vspace{-\baselineskip}

In order to fully treat the uncertainty, we need a probe of the systematic uncertainty produced by the ionosphere. As a proxy for this, we choose to measure the scatter of each of our calibrators relative to each of their mean positions across all epochs. We then measure the covariance of this distribution and add it to the simulated positional covariance at each epoch. The scale of this systematic uncertainty is approximately 1 mas (found from the diagonal elements in the covariance matrix), and is only very mildly correlated in R.A. and Declination. The uncertainties reported in Table \ref{table: positions} include this systematic atmospheric term.

\subsection{Parallax and Proper Motion Fitting} \label{sec: mcmc}

For fitting parallax and proper motion we adopt the model used by Gaia as described in \citet{2016A&A...595A...4L} Equation 2, similarly assuming that the effect of radial proper motion is small enough as to be undetectable in our observations. Using that equation, a position can be found as a function of time, given the set of variables \{$\alpha_0$, $\delta_0$, $\omega$, $\mu_\alpha^*$, $\mu_\delta$\} and choosing a reference epoch. For our analysis we choose a reference epoch of MJD 58849, which corresponds to January 1st 2020, slightly prior to our first observation.

We implement a Gaussian prior on the total proper motion of $115\pm33$ mas yr$^{-1}$, as reported in Paper I, though we note that the final parameter estimates do not seem to be strongly effected by this prior. Because our position uncertainties are covariant, the equation for log-likelihood takes on the form

\begin{equation}
    \ln{\mathcal{L}} = - \frac{1}{2} \sum_{i=0}^n \left[ 2 \ln{2\pi} + \ln{(\det{V_i})} + x_i^T\ V_i^{-1} x_i \right] \mathrm{,}
\end{equation}

\noindent where at a given epoch $i$, $x_i=(m_i-o_i)$ represents the difference between the model position $m_i=(\alpha_m, \delta_m)_i$ and observed position $o_i=(\alpha_o, \delta_o)_i$, and $V_i$ represents the observed/estimated covariance. We then pass the data, uncertainties, prior, and log-likelihood functions to the \textit{emcee} Python package. We use 200 walkers initialized in a small space around initial approximate estimates provided by least-squares fitting of a straight line, and the scatter relative to that. The walkers progress for 20,000 steps, after which we calculate the maximum auto-correlation time $\tau$ (typically of order 50 steps) and use that to discard the first $3\tau$ steps (prior to burn-in) and thin by a factor of $\tau/2$. These sample chains are flattened, and then can be evaluated to find best fit values and uncertainties for each of our parameters as shown in Table \ref{table: parameters}.

\begin{deluxetable}{lc} \label{table: parameters}
\tablecaption{Estimated and Calculated Values}
\tablehead{\colhead{Parameter} & \colhead{Estimate}}
    \startdata
    $\alpha_0$ & 00h02m58.20346s $\pm\ 0.89$ mas \\
    $\delta_0$ & +62d16m09.51294s $\pm\ 0.81$ mas \\
    $\omega$ & $0.63 \pm 0.45$ mas \\
    $\mu_\alpha^*$ & $32.52 \pm 0.59$ mas yr$^{-1}$ \\
    $\mu_\delta$  & $-13.71 \pm 0.53$ mas yr$^{-1}$ \\
    \tableline
    $\mu_{\text{tot}}$  & $35.30 \pm 0.60$ mas yr$^{-1}$ \\
    $\theta_{\text{PA}}$ & $112.86 \pm 0.83$ deg \\
    Age & $47.60 \pm 0.80$ kyr \\
    $v_{\text{2kpc}}$\tablenotemark{a}  & $334.90 \pm 5.66$ km s$^{-1}$ \\
    $v_{\text{3kpc}}$\tablenotemark{a}  & $502.35 \pm 8.4$ km s$^{-1}$ \\
    \enddata
\tablecomments{Values provided are the median of all samples for each parameter. Uncertainties are reported at the 1-$\sigma$ level.}
\tablenotetext{a}{These velocities assume distances with no uncertainty.}
\end{deluxetable}

\vspace{-1\baselineskip}

\section{Results} \label{sec: results}

\begin{figure*}
    \centering
    \includegraphics[width=\linewidth]{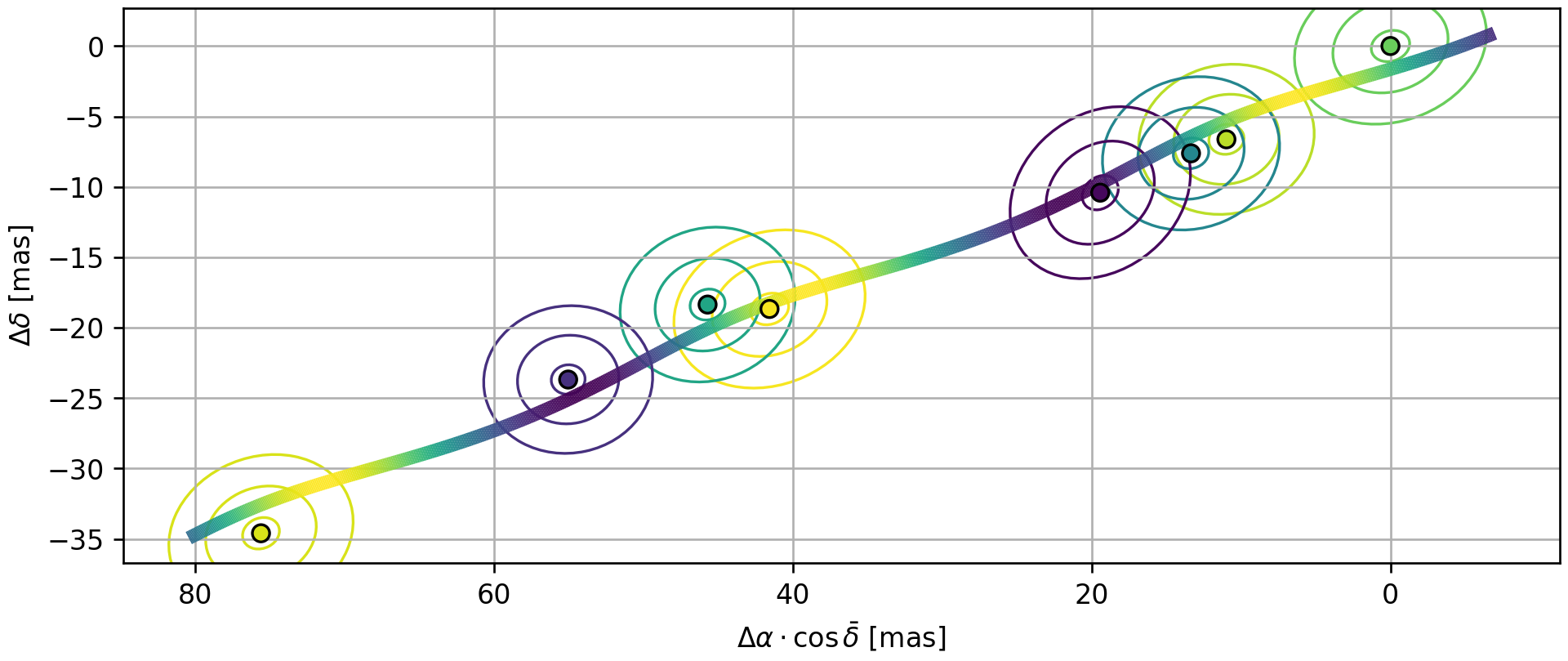}
    \caption{The datapoints along with the median fit produced by our program. Here, color corresponds to time, such that purple (the beginning of the viridis scale) represents the beginning of the year. The closer a data point is to a similar color in the fit, the better the fit. Ellipses surrounding each data point represent the 1, 2, and 3$\sigma$ positional uncertainties.}
    \label{fig: median}
\end{figure*}

\begin{figure}
    \centering
    \includegraphics[width=\linewidth]{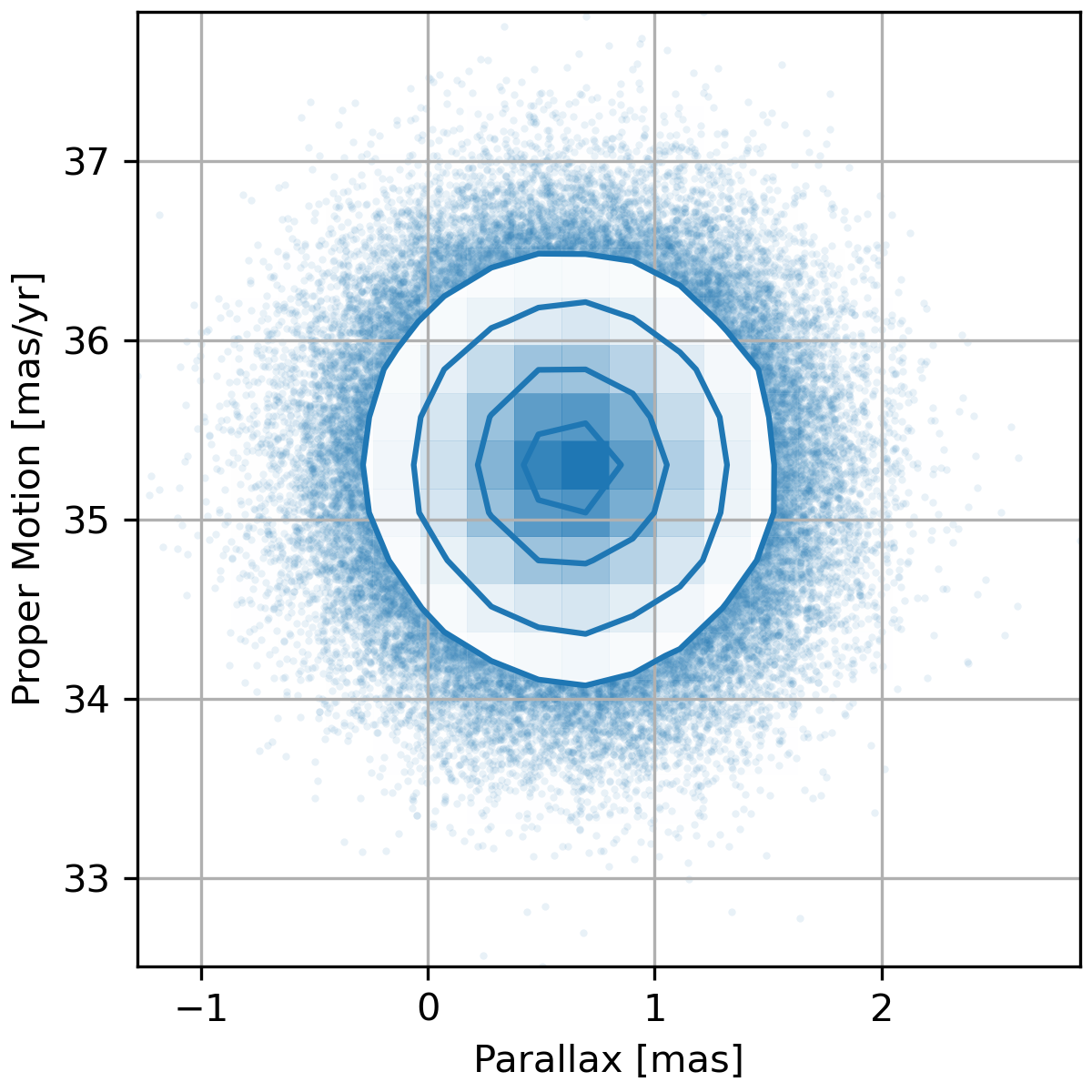}
    \caption{A 2D histogram of the relevant parameter distributions. Here we show the total proper motion, calculated using its individual components.}
    \label{fig: blob}
\end{figure}

Our best-fit model is illustrated in Figure \ref{fig: median}. Here, we mainly focus on the estimates of parallax and proper motion, as those have the largest physical implications. From the best-fit values for proper motion along each axis, we can calculate the total proper motion magnitude $\mu_\mathrm{tot} = \sqrt{\mu_\alpha^{*2} + \mu_\delta^2} = 35.30 \pm 0.60$ mas yr$^{-1}$ and position angle $\theta_\mathrm{PA} = \arctan{\left({\mu_\alpha^*}/{\mu_\delta}\right)} = 112.86^\circ\pm0.83^\circ$. As the total proper motion is generally of more interest than its individual components, we illustrate the distribution in this parameter against parallax in Figure \ref{fig: blob}. As the full five-parameter corner plot is quite large and does not show any correlations of note, we opt to make it available upon request.

If we compare our total proper motion and position angle with the initial estimates of these parameters found from \textit{Fermi} data in Paper I ($\mu_\mathrm{tot} = 115\pm33$ mas yr$^{-1}$ and $\theta_\mathrm{PA} = 121^\circ\pm13^\circ$), we find differences of $2.37\sigma$ and $0.59\sigma$, respectively. The agreement in position angle is expected, as the prior estimate of that value also matched well with the direction of the orientation of the pulsar wind nebula as described in Paper I and later in \citet{2023ApJ...945..129K}, \textbf{with the later finding a position angle of $\theta_\mathrm{PA} = 111.13^\circ\pm0.52^\circ$}. 

Next we examine the parallax of the source, for which a value of $\omega=0.63$ can be naively inverted to find a distance and velocity of 1.59 kpc and 244 km s$^{-1}$ respectively. That said, such a naive inversion will suffer from the Lutz-Kelker bias, and can produce meaningless or aphysical values in cases where the parallax is not incredibly well constrained \citep{2010MNRAS.405..564V}. Ultimately a proper measurement of the distance to J0002+6216 via parallax is thwarted by the effects of the ionosphere. The scale of its perturbation on our observations is above the scale we expect to see the parallax at, and in fact Paper I showed that we can largely rule out parallaxes larger than 1 mas or so. 

Perhaps the best estimate for distance currently possible comes from the association with the remnant CTB-1. The new estimates of proper motion trace back a path directly through the geometric center of the SNR (as seen in Figure \ref{fig: path} and discussed more in Section \ref{sec: discussion}), making a chance association quite unlikely. This geometric argument has been sufficient for other systems such as the recently discovered "Mini Mouse" \citep{2023MNRAS.523.2850M}. If we assume this association to be true, then our best constraints for the distance to the pulsar are once again approximately 1.5-4 kpc \citep{1982AJ.....87.1379L,1994ApJ...434..635H,1997AJ....113..767F,2004ApJ...616..247Y}, plus an upper limit $d\lesssim7$ kpc derived from X-ray and $\gamma$-ray efficiencies \citep{2018MNRAS.476.2177Z,2018ApJ...854...99W}.

\section{Discussion} \label{sec: discussion}

\begin{figure*}
    \centering
    \includegraphics[width=\linewidth]{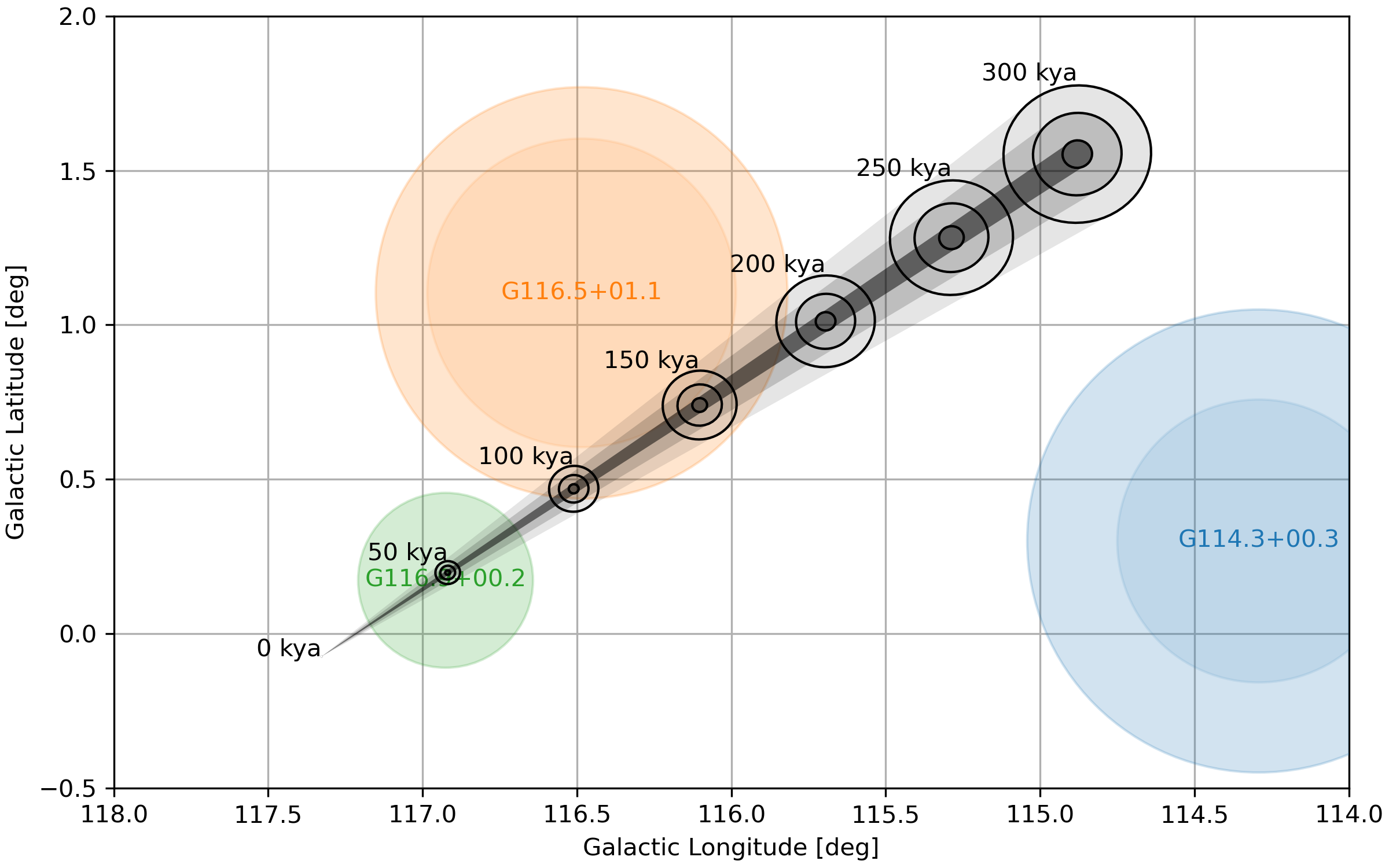}
    \caption{The possible history of the pulsar, accounting for uncertainties in position and proper motion. While \psr\ is likely to have originated at or near the geometric center of \ctb\ (\gctb\ in this Figure), here we illustrate the possible positions if we assume that it originated somewhere beyond the SNR, out to approximately the characteristic age of the pulsar. The black circles represent that possible $1\sigma$, $3\sigma$, and $5\sigma$ positional scatter at each epoch, whereas the shaded regions represent the paths between those epoch contours. Colored circles represent remnants found in \citet{2019JApA...40...36G}.}
    \label{fig: path}
\end{figure*}

Given our significantly improved measurement of the proper motion, we can begin to discuss the implications of the pulsar's space-velocity. This velocity can now be written as 

\begin{equation}
    v = 334.90 \left(\frac{d}{2\ \mathrm{kpc}}\right) \mathrm{km\ s}^{-1} \text{,}
\end{equation}

\noindent where we have chosen to use 2 kpc as our reference point to match the assumptions in Paper I. Our new estimate has the effect of decreasing the speed by a factor of more than three from the original estimate of over 1000\,km\,s$^{-1}$. For the range 1.5--4\,kpc we find velocities 251--670\,km\,s$^{-1}$, and more specifically for the value of $3.1\pm0.4$\,kpc obtained by \citet{1994ApJ...434..635H} we find $519.09\pm67.56$ km\,s$^{-1}$. While any of these values would still place \psr\ among a rare category of high-velocity pulsars compared to the overall distribution \citep{1997MNRAS.291..569H,2017A&A...608A..57V}, it could perhaps now be considered for demotion from the ranks of other Cannonball pulsars \citep[see][]{2005MNRAS.360..974H,2019ApJ...877...78S} to something more akin to a vigorously thrown bowling ball.

Next, we examine the implications of our proper motion result for \psr, using three different measures of the pulsar age. This approach does not require estimates of other (uncertain) parameters such as the distance, and thus should yield secure results. In the first instance, we assume that the age of \psr\ is equal to its characteristic age ($\tau_c$=306 kyrs). In this case, our measured proper motion places the pulsar birthplace more than $2^\circ$ away, well outside the boundaries of \ctb. In this instance, there is no physical PSR-SNR association, just a line-of-sight coincidence. There are at least two weaknesses with this hypothesis. \citet{2012ApJ...746..105N} have shown that long PWN tails that point back toward SNRs have a very low \textit{posterior} probability of occurrence. If anything, the chance probability is even smaller for this system since both our proper motion result and the PWN tail of \psr\ can be accurately traced back to the geometric center of \ctb. 

Another objection could be making the (common) assumption that $\tau_c$ can be used as a proxy for the actual age of the system. This has been shown to have questionable accuracy, at least for young pulsars in which $\tau_c$ is seen to be substantially lower than or greater than the quoted age of the associated SNR \citep{2006ApJ...652.1523B, 2012Ap&SS.341..457P, 2021MNRAS.507L..41J}. This discrepancy with characteristic age implies a relatively long birth spin period, close to the present period of 115 ms, and has marked implications for kick models which predict a close association between the direction of proper motion and the rotation axis \citep[e.g.][]{2007ApJ...660.1357N, 2022MNRAS.517.3938C}. Examining this alignment would require more sensitive and full Stokes measurement of the pulsar profile to improve existing measurements of both the rotation and magnetic axes relative to the line of sight \citep{2018ApJ...854...99W}. Figure \ref{fig: path} illustrates the path of the pulsar over time, out to the distance that would be covered under this assumption of age.

Motivated by these findings, we consider the implications if the pulsar age is equal to the age of the SNR of 10$\pm$2 kyrs, derived in Paper I from several sources in the literature. Using the measured proper motion, we find that the pulsar will have traveled only 6$^\prime$ from current location, placing it \textit{outside} the southeastern edge of \ctb\ and thus is not associated. However, the difficulty in adopting this age is that we would need to explain how \psr\ was born inside the tail of its own PWN. In our original discovery image (Paper I) the tail is at least 7$^\prime$ in length, and in the lower resolution Canadian Galactic Plane Survey images \citep[CGPS,][]{2006A&A...457.1081K} it can be traced 11$^\prime$ back to the edge of \ctb. It seems more likely that the true age of \psr\ is greater than 10 kyr.

Finally, we can examine the kinematic age of the system. If we assume that the pulsar originated from the geometric center of the remnant, then we can use the 28$^\prime\pm{1}^\prime$ offset between the pulsar and the geometric center to find a kinematic age for a system of approximately $47.6\pm0.8$ kyr. This has the effect of implying \ctb\ is significantly older than the estimates discussed above. However, this is not completely unexpected, as there is substantial evidence that \ctb\ is in its radiative phase, and more specifically belongs to a class of mixed morphology remnants, as evidenced by its ring-like radio emission and centrally located X-ray emission \citep{2006ApJ...647..350L}. This means that typical Sedov-Taylor scaling relationships \citep{1959sdmm.book.....S,1950RSPSA.201..159T} one might use for younger remnants may not produce accurate results from X-ray properties and could vary significantly depending on various assumptions. 

To illustrate the effect of this increased age on the parameters of the system, we can model the approximate evolution of \ctb\ over time, following the procedure roughly described in \citet{2012A&ARv..20...49V}. The remnant initially follows the Sedov-Taylor equations up until radiative losses become important, and further evolution is directed by momentum conservation. Adopting the equation for the latter phase from \cite{2009MNRAS.395..351T}, our model can be written as

\begin{subequations}
    \begin{equation}
        R(t) \approx
        \begin{cases}
        5.03\ ({E_{51}}/{n_0})^{1/5}\ t^{2/5}\ \mathrm{pc}, & t < t_{rad} \\ 
        r_{rad} \left( 1.58\ \frac{t}{t_{rad}} - 0.58 \right)^{1/4}, & t \geq t_{rad} 
        \end{cases}
    \end{equation}
    \begin{equation}
        t_{rad}\approx 44.6\ ({E_{51}}/{n_0})^{1/3}\ \mathrm{kyr}
    \end{equation}
    \begin{equation}
        r_{rad} \approx 23\ ({E_{51}}/{n_0})^{1/3}\ \mathrm{pc}
    \end{equation}
\end{subequations}

\noindent where $t_{rad}$ and $r_{rad}$ are the time and radius at which the remnant transitions into the radiative phase. Note that this evolution is dependent on the term $({E_{51}}/{n_0})$, the ratio of the supernova energy (in units of $10^{51}$ ergs) and the pre-shock hydrogen density (in units of cm$^{-3}$). This means that for a given choice of energy, density and distance, one can accurately describe the angular radius as a function of time, allowing constraints on those parameters knowing the current angular size of \ctb\ ($\sim 17.5^\prime$) and its presumed age ($\sim48$ kyr). From this we can see that from reasonable choices of distance, $({E_{51}}/{n_0}) \approx$ 0.01--0.5, which implies a high-density medium, a weak supernova, or some combination of the two. 

\begin{figure}
    \centering
    \includegraphics[width=\linewidth]{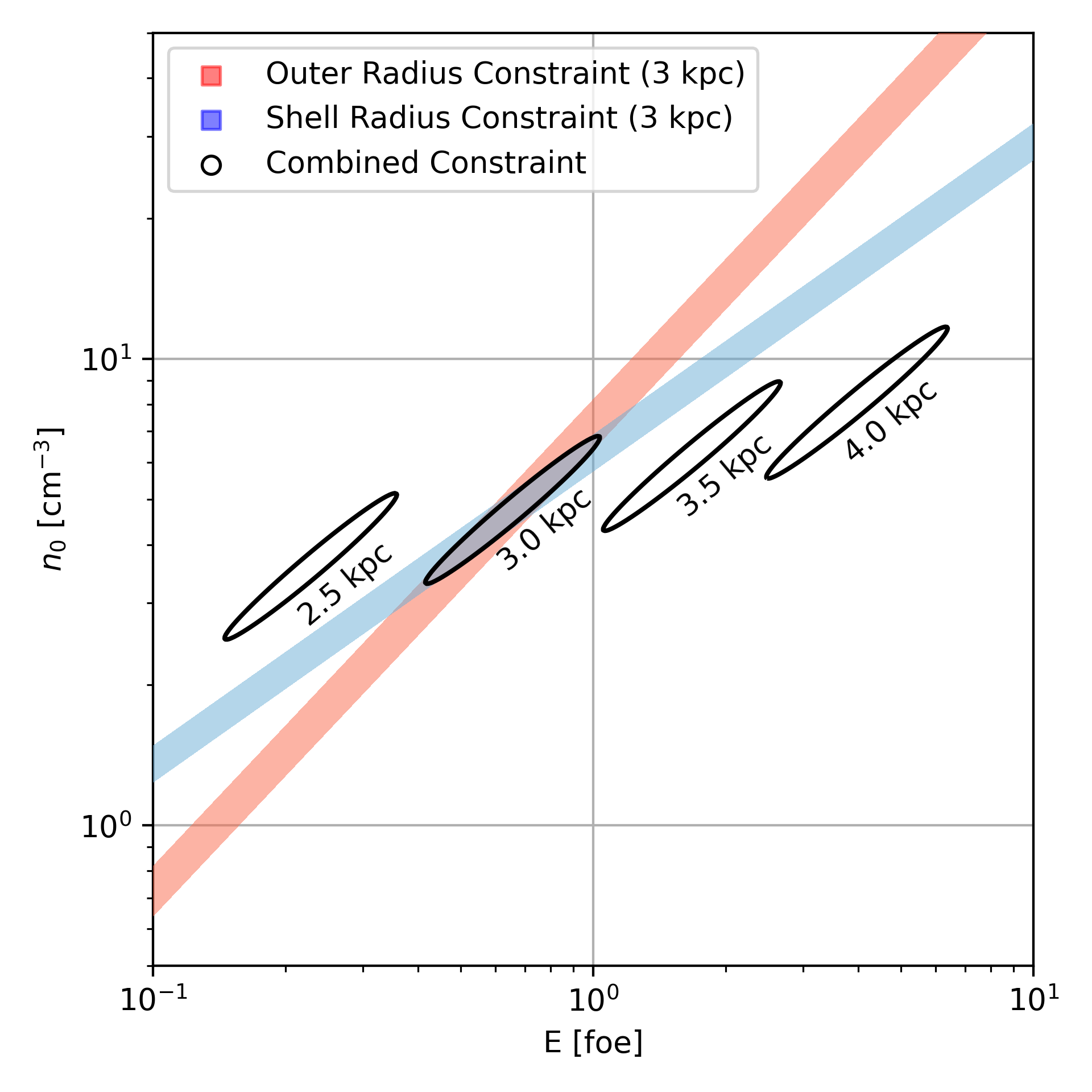}
    \caption{Constraints in the density/energy space derived from the observed outer and inner radii of \ctb, assuming a distance of 3 kpc \citep[near that found in][]{1994ApJ...434..635H}. The red region represents outer radii which are within $1^\prime$ of $17.5^\prime$, while the blue region represents inner/shell radii which are within $2^\prime$ of $13^\prime$. The space where both regions overlap is shown for 3 kpc as well as a few other distances.}
    \label{fig: outerinner}
\end{figure}

Furthermore, for mixed morphology remnants \citet{1999ApJ...524..179C} derived that the time/radius at which the radio shell will begin to form can be written as

\begin{equation}
    t_{shell} \approx 53\ E_{51}^{3/14} n_0^{-4/7}\ \mathrm{kyr}
\end{equation}
\begin{equation}
    r_{shell} \approx 24.9\ E_{51}^{2/7} n_0^{-3/7}\ \mathrm{pc}
\end{equation}

\noindent allowing one to place constraints similarly using the inner edge of the radio emission ring (which has an angular size of $\sim 13^\prime$). Figure \ref{fig: outerinner} shows such constraints for the inner and outer radii, assuming a distance of 3 kpc so as to probe the supernova energy and local pre-shock density. Most choices of distance seem to require fairly typical energies, and the high value for density could be expected given that most mixed-morphology remnants seem to form in regions of denser than typical ISM. At the 3 kpc distance, the best match to the constraints comes when $E_{51}\approx0.7$ and $n_0\approx4$ cm$^{-3}$.

Interestingly, this result of a higher than expected density also has implications for the bow shock of \psr. As described in \citet{2023ApJ...945..129K}, the pulsar is moving rapidly through the local ISM, generating a bow shock around it and trailing a long and highly collimated tail of emission for several arcminutes. Paper I estimated a Mach number $\mathcal{M} \approx 200$ from initial measurements and comparisons with systems with similar morphology \citep{2012ApJ...746..105N,2008AIPC..983..171K}. Assuming the distance and density used as an example above (and a 10\% helium mixture as in Paper I), we estimate that the pulsar would have a mach number $\mathcal{M}\approx 110$, still broadly consistent with the observed collimation of the tail. Without an increase in density, the decrease in velocity due to the updated proper motion would lower the overall Mach number, resulting in a less striking collimation. Change in various parameters also has the effect of lowering the standoff distance of the shock, derived by equating the wind pressure and the ram pressure of the pulsar \citep{1996ApJ...464L.165F} to

\begin{equation}
    r_{SO} = 0.0015\ \left( \frac{n_0}{3.5\ \mathrm{cm}^{-3}} \right)^{-1/2} \left( \frac{d}{3\ \mathrm{kpc}} \right)^{-1} \mathrm{pc}\text{,}
\end{equation}

\noindent which would be a few times smaller than the best estimates from \citet{2023ApJ...945..129K}, and among the smallest standoffs known.

\section{Conclusions} \label{sec: conclusions}

Using several years worth of HSA data, we have significantly refined the estimate of the proper motion of the ``Cannonball'' pulsar J0002+6216, to a new value of $\mu_\mathrm{tot} = 35.30 \pm 0.60$ mas yr$^{-1}$. This new estimate is more than three times smaller than the previous best estimates made in \citet{2019ApJ...876L..17S} using \textit{Fermi} data, and as such would act to lower the spatial velocity perpendicular to the line of sight. This new proper motion also provides for a precise distance-independent kinematic age measurement for the system of $47.60\pm0.80$ kyr, significantly older than previously estimated from the pulsar/remnant, but also still significantly younger than the pulsar's characteristic age.

We also attempt to provide a more accurate measure of the distance to the source, as that will also play a role in determining the kinematic properties. While a value for the parallax of $\omega = 0.63 \pm 0.45$ mas was obtained, the high uncertainties make a definitive distance difficult to determine. These uncertainties are dominated by scattering in the ionosphere, but we also see uncertainty simply from trying to fit a position to a fairly weak source. Refinement of this value would require both increased sensitivity and more complex observing strategies, which could be used to isolate and remove these ionospheric effects. We note that this is likely to be an excellent target for the ngVLA in the coming decade, as it will have baselines similar to the present-day VLBA, but with significantly more antennas allowing for a dramatic increase in sensitivity. 

As an example, in one hour the HSA\footnote{\url{http://old.evlbi.org/cgi-bin/EVNcalc.pl}} can be expected to reach a thermal noise level of approximately 13 uJy at 1.4 GHz, which then necessitates longer and more difficult to schedule observations to provide enough signal to noise for a 14 uJy source such as \psr. In comparison, a one hour observation on the ngVLA\footnote{\url{https://ngvla.nrao.edu/page/performance}} at 2.4 GHz should reach a thermal noise of 0.24 uJy. Even if we assume that the pulsar will dim at this higher frequency by a factor of about 2, following the standard pulsar spectral index of $\alpha\approx-1.4$ \citep{2013MNRAS.431.1352B}, we should still expect an approximate signal-to-noise ratio of 27 after just one hour. Combine this with the possibility of more advanced observation and calibration techniques (such as simultaneous subarray observations of a calibrator source), and it is likely that ngVLA could provide an estimate of the parallax at significantly reduced uncertainty. 

Beyond the parallax, further questions also remain for this system. The proper motion, now reduced in magnitude but with substantially less uncertainty, combined with the expansion of \ctb\, seem to be indicative of a fairly large ISM density. The implied value would be beyond the low density ($\sim$0.1--1.0 cm$^{-3}$) which has typically been inferred from the remnant's relatively weak H$\alpha$ emission \citep{1982AJ.....87.1379L} or from the non-detection of tracers such as CO \citep{2023ApJS..268...61Z}. While this can be somewhat mitigated if one assumes a weaker supernova energy ($E_{51}<1$), the collimation of the pulsar's bow shock is decoupled from this term and still requires a higher density given this lower velocity. Further work will be necessary to properly constrain the excess density and identify its origin.

\section{Acknowledgements} \label{sec: acknowledgments}

SB, FKS, and GBT acknowledge support from NASA Fermi Guest Investigator Program, grants 80NSSC18K1725, 80NSSC19K1508, 80NSSC22K1643.  Work at NRL is supported by NASA.  The National Radio Astronomy Observatory is a facility of the National Science Foundation operated under cooperative agreement by Associated Universities, Inc. Support for this work was provided by the NSF through the Grote Reber Fellowship Program administered by Associated Universities, Inc. National Radio Astronomy Observatory and the Student Observing Support Program. We thank Tom Maccarone, Adam Deller, and David Green for insightful discussions toward the goals of this work, as well as Jason Wu for his supporting of early HSA observations from Effelsberg.

This work has made use of data from the European Space Agency (ESA) mission {\it Gaia} (\url{https://www.cosmos.esa.int/gaia}), processed by the {\it Gaia} Data Processing and Analysis Consortium (DPAC, \url{https://www.cosmos.esa.int/web/gaia/dpac/consortium}). Funding for the DPAC has been provided by national institutions, in particular the institutions participating in the {\it Gaia} Multilateral Agreement.

This research has made use of NASA’s Astrophysics Data System and has made use of the NASA/IPAC Extragalactic Database (NED) which is operated by the Jet Propulsion Laboratory, California Institute of Technology, under contract with the National Aeronautics and Space Administration. This research has made use of data, software and/or web tools obtained from NASA’s High Energy Astrophysics Science Archive Research Center (HEASARC), a service of Goddard Space Flight Center and the Smithsonian Astrophysical Observatory, of the SIMBAD database, operated at CDS, Strasbourg, France.

\software{\href{www.astropy.org}{Astropy} \citep{2022ApJ...935..167A},
          \href{https://casa.nrao.edu/}{CASA} \citep{2022PASP..134k4501C},
          \href{https://corner.readthedocs.io/en/latest/}{corner.py} \citep{2016JOSS....1...24F},
          \href{https://www.cv.nrao.edu/adass/adassVI/shepherdm.html}{Difmap} \citep{1997ASPC..125...77S},
          \href{https://emcee.readthedocs.io}{emcee} \citep{2013PASP..125..306F}
          \href{www.matplotlib.org}{matplotlib} \citep{MatplotlibCitation},
          \href{www.numpy.org}{Numpy} \citep{NumpyCitation},
          \href{www.scipy.org}{Scipy} \citep{ScipyCitation}}

\bibliography{refs,refs2}{}
\bibliographystyle{aasjournal}

\end{document}